\documentclass[superscriptaddress,twocolumn]{revtex4}
\usepackage{amssymb}
\usepackage[tbtags]{amsmath}
\usepackage{graphicx}
\usepackage{epsfig,graphicx,times}

\begin{document}

\title{Testing Bell inequalities with circuit QEDs by joint spectral measurements}
\author{Hao Yuan}
\affiliation{Quantum Optoelectronics Laboratory, School of Physics
and Technology, Southwest Jiaotong University, Chengdu 610031,
China}\affiliation{Clarendon Laboratory, University of Oxford, Parks
Road, Oxford OX1 3PU, United Kingdom}
\author{L. F. Wei}
\email{weilianfu@gmail.com}\affiliation{Quantum Optoelectronics
Laboratory, School of Physics and Technology, Southwest Jiaotong
University, Chengdu 610031, China}\affiliation{State Key Laboratory
of Optoelectronic Materials and Technologies, School of Physics and
Engineering, Sun Yat-Sen University, Guangzhou 510275, China}
\author{J. S. Huang}
\affiliation{Quantum Optoelectronics Laboratory, School of Physics
and Technology, Southwest Jiaotong University, Chengdu 610031,
China}
\author{X. H. Wang}
\affiliation{Quantum Optoelectronics Laboratory, School of Physics
and Technology, Southwest Jiaotong University, Chengdu 610031,
China}
\author{Vlatko Vedral}
\email{vlatko.vedral@qubit.org} \affiliation{Clarendon Laboratory,
University of Oxford, Parks Road, Oxford OX1 3PU, United
Kingdom}\affiliation{Centre for Quantum Technologies, National
University of Singapore, 3 Science Drive 2, Singapore 117543,
Singapore}\affiliation{Department of Physics, National University of
Singapore, 2 Science Drive 3, Singapore 117542, Singapore}

\date{\today}

\begin{abstract}
We propose a feasible approach to test Bell's inequality with the
experimentally-demonstrated circuit QED system, consisting of two
well-separated superconducting charge qubits (SCQs) dispersively
coupled to a common one-dimensional transmission line resonator
(TLR). Our proposal is based on the joint spectral measurements of
the two SCQs, i.e., their quantum states in the computational basis
$\{|kl\rangle,\,k,l=0,1\}$ can be measured by detecting the
transmission spectra of the driven TLR: each peak marks one of the
computational basis and its relative height corresponds to the
probability superposed. With these joint spectral measurements, the
generated Bell states of the two SCQs can be robustly confirmed
without the standard tomographic technique. Furthermore, the
statistical nonlocal-correlations between these two distant qubits
can be directly read out by the joint spectral measurements, and
consequently the Bell's inequality can be tested by sequentially
measuring the relevant correlations related to the suitably-selected
sets of the classical local variables $\{\theta_j,\theta_j',
j=1,2\}$. The experimental challenges of our proposal are also
analyzed.
\vspace{0.3cm}

PACS number(s):
 03.65.Ud, 
 42.50.Dv, 
 85.25.Cp  

\end{abstract}
\maketitle

\section{Introduction}

Historically, the well-known Einstein, Podolsky, and Rosen (EPR)
paradox~\cite{EPR1935} concerning the completeness of quantum
mechanics was proposed based on a gedanken experiment in 1935. EPR
claimed that quantum mechanics is incomplete and so-called local
hidden variables (LHV) should exist. Subsequently, it has provoked
much debate on the completeness of quantum mechanics and the
existence of LHV theories. In 1964, Bell actually quantified the EPR
argument and derived a strict inequality~\cite{Bell1964} with
respect to the correlation strengths possible to achieve by all the
LHV models. If this inequality is violated, then there are no LHV
and the quantum mechanical prediction of the existence of quantum
nonlocal correlation (i.e., entanglement) is sustained. However,
Bell's seminal inequality did not allow for the practically-existing
imperfections and thus was not accessible to the experimental tests.
Soon later, Clauser, Horne, Shimony and Holt (CHSH) addressed this
issue and derived an experimentally testable
inequality~\cite{CHSH1969}.
During the past decades, a number of experimental tests of CHSH-type
Bell's inequality have been reported using, e.g., entangled photon
pairs~\cite{photon pairs}, trapped ions~\cite{Wineland2001}, an atom
and a photon~\cite{Monroe2004}, two superconducting phase
qubits~\cite{Martinis2009}, and even two entangled degrees of
freedom (comprising spatial and spin components) of single
neutrons~\cite{singleneutron}, and so on. These experimental
evidences strongly convince that Bell's inequality could be violated
and thus agree well with quantum mechanical predictions, ruling out
the LHV theories.

Recently, a novel superconducting electrical circuit architecture
(called circuit QED) analogous to cavity QED has been first
suggested by Blais et al.~\cite{Blais2004}, and then realized in the
experiment by Wallraff et al.\cite{Wallraff2004}. In circuit QED, a
superconducting qubit is strongly coupled to one-dimensional
transmission line resonator (TLR) which acts as quantized cavity.
Now much attention has been focused on this field because it opens
the possibility of studying quantum optics phenomena in solid-state
system and quantum information processing. Experimentally,
remarkable advances have been achieved such as: observation of the
vacuum Rabi splittings~\cite{Wallraff2004} and AC Stack
shifts~\cite{Schuster2005}, observing Berry's phase~\cite{Leek2007},
measuring the Lamb shifts~\cite{lambshift2008}, generations of
microwave single photons~\cite{Houck07} and Fock
states~\cite{Hofheinz08}, coupling two qubits via cavity as quantum
bus~\cite{Majer07,Martinis2009}, two-qubit Grover and Deutsch-Jozsa
algorithms~\cite{DiCarlo09}, etc.
In this system the qubit-state readouts were achieved by detecting
the state-dependent shifts of the frequency of the
dispersively-coupled resonator~\cite{DiCarlo09, FilippPRL102,
ChowPRA81, ReedPRL105, BianchettiPRA80}. Indeed, in the dispersive
regime~\cite{Schuster07}, the transition frequencies of the
superconducting qubits are far detuned from the frequency of the
coupled resonator. As a consequence, a sufficiently large
state-dependent frequency shifts of the resonator can be
induced~\cite{Blais2004,Wallraff2004}, which could be tested by
measuring the transmission spectra of the driven resonator.

Very recently, an efficient approach to implement the joint spectral
measurements of the multi-qubits has been proposed by detecting the
transmission spectra of the common resonator~\cite{Wei}. By
considering all the quantum correlation of qubits and resonator
(i.e., beyond the usual coarse-grained/mean-field approximation), it
is shown that each peak in the measured transmission spectra of the
driven resonator marks one of the logic states and the relative
height of such a peak is related to its corresponding superposed
probability. With these joint spectral measurements, Huang et
al.,~\cite{Huang} presented a high efficiency tomographic
reconstructions of more than one-qubit quantum states.
As another application, in this paper we discuss how the proposed
joint spectral measurements can be further utilized to test Bell's
inequality with an experimentally-demonstrated two-qubit circuit QED
system. Compared with the previous schemes for testing Bell
inequalities, the advantage of the present proposal is that, the
desirable coincided measurements on the two qubits can be directly
implemented by the suitably-designed joint spectral measurements.
This is because the states of the two qubits jointly determine the
frequency shifts and consequently the spectral distributions of the
driven resonator. Therefore, the nonlocal correlations between the
well-separated qubits can be directly read out by analyzing the
measured transmission spectra of the driven resonator. We note in
passing that none of the circuit QED proposals so far can be used to
close the locality loophole.

The paper is organized as follows: In section II, we propose a
method to deterministically generate an ideal Bell state with the
experimentally-demonstrated circuit QED system, via an {\it i}SWAP
gate assisted by several single-qubit gates~\cite{Blais07}. Further,
a simple quantum interference method, instead of the usual
quantum-state tomographic reconstruction, is proposed to confirm
such a preparation by coherent quantum operations and projective
measurements on two qubits in the computational basis. In section
III, we discuss how to utilize the joint spectral measurements to
implement this confirmation. Importantly, by applying two individual
Hadamard-like operations to encode the local variables
$\{\theta_1,\theta_2\}$ into the generated Bell state, we show that
the statistical correlations between the distant qubits can be read
out by these spectral measurements for various typical sets of
classical local variables $\{\theta_j,\theta_j', j=1,2\}$. With
these measured correlations the CHSH-type Bell's inequality can be
tested. Discussion and conclusion are presented in section IV.

\section{Generation and confirmation of Bell states}

In this section, we will show how to deterministically prepare one
of the ideal Bell states
\begin{eqnarray}
|\Phi_\pm\rangle=\frac{1}{\sqrt{2}}(|00\rangle\pm|11\rangle),
|\Psi_\pm\rangle=\frac{1}{\sqrt{2}}(|01\rangle\pm|10\rangle),
\end{eqnarray}
with the experimentally-demonstrated circuit QED system. In fact,
the Bell state in this system have been prepared by several methods,
including measuring-based synthesis~\cite{Blais09} and gate sequence
(see, eg. Refs.~\cite{FilippPRL102,ChowPRA81,Majer07,Martinis2009,
Blais07}). Here, we will present an alternative gate sequence method
to implement the desirable preparation based on an {\it i}SWAP gate
(the previous methods use the controlled phase gate).

\begin{figure}[htbp]
  \centering
  \includegraphics[width=7cm]{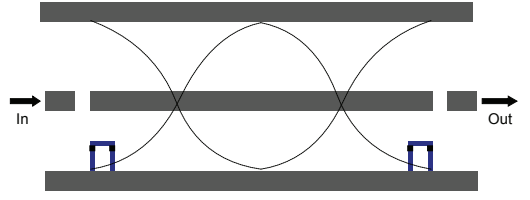}\\
  \caption {(Color online) Circuit QED architecture with two superconducting charge
  qubits (SCQs) dispersively coupling to one-dimensional transmission line
  resonator (TLR). The joint spectral measurements of the two SCQs can be realized by
  detecting the state-dependent frequency shifts in the transmission spectra of
  the driven resonator.}\label{Fig1}
\end{figure}

We consider the circuit QED architecture sketched in Fig.1, in which
two superconducting charge qubits (SCQs) are coupled to
one-dimensional TLR. To obtain the maximal coupling, two SCQs are
placed close to the ends of the resonator (voltage antimodes of the
resonator). When two SCQs work at their optimal point and under the
rotating-wave approximation (RWA), the system of two SCQs plus
resonator is described by the usual Dicke Hamiltonian
($\hbar=1$)~\cite{Blais07}
\begin{eqnarray}
H=\omega_ra^\dagger
a+\sum_{j=1,2}[\frac{\omega_j}{2}\sigma_{zj}+g_j(a^\dagger
\sigma_{-j}+a\sigma_{+j})].
\end{eqnarray}
Here, $\omega_r$ is the single-mode frequency of the TLR,
$a^{\dagger}(a)$ its creation (annihilation) operator; $\omega_j$
the transition frequency of $j$th SCQ that can be adjusted by
external bias flux and $g_j$ the coupling strength of the $j$th SCQ
to the resonator. Finally, $\sigma_{+j}=|1\rangle_j\langle 0|$,
$\sigma_{-j}=|0\rangle_j\langle 1|$ and
$\sigma_{zj}=|0\rangle_j\langle 0|-|1\rangle_j\langle1|$.
Two SCQs in this circuit can be controlled and measured by driving
the resonator. The Hamiltonian to describe such a drive is given by
\begin{eqnarray}
H_d=\epsilon(a^{\dagger}e^{-i\omega_dt}+ae^{i\omega_dt}),
\end{eqnarray}
where $\epsilon$ is time-independent real amplitude and $\omega_d$
the frequency of the externally applied drive.

From the easily-prepared initial state $|00\rangle_{12}$, an ideal
Bell state can be generated deterministically by the following three
steps, the corresponding gate sequence is shown in Fig.~2.

\begin{figure}[htbp]
  \centering
  \includegraphics[width=7cm]{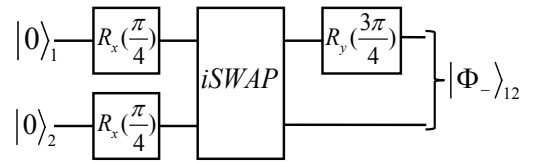}\\
  \caption {Gate sequence to generate the Bell state $|\Phi_-\rangle_{12}$
  from the initial state $|00\rangle_{12}$.}\label{Fig2}
\end{figure}

{\it Step 1}: Perform the unitary operation
$R_{x}(\pi/4)=\exp(i\sigma_{x}\pi/4)$ on each SCQ to prepare the
superposition of two basis states of each SCQ. That is,
\begin{eqnarray}
|00\rangle_{12}\stackrel{R_{x1}(\frac{\pi}{4})R_{x2}(\frac{\pi}{4})}{\rightarrow}|\psi\rangle_{12}
=\frac{1}{2}(|0\rangle+i|1\rangle)_1(|0\rangle+i|1\rangle)_{2}.
\end{eqnarray}
The unitary operations $R_{x1}(\pi/4)$ and $R_{x2}(\pi/4)$ can be
achieved by externally applied microwave drives with the duration of
$t_1=-\pi\Delta_r/4\epsilon g_1$ and $t_2=-\pi\Delta_r/4\epsilon
g_2$, respectively~\cite{Blais07}. Here,
$\Delta_{r}=\omega_d-\omega_r$ is the detuning between the drive
frequency and the resonator one.

{\it Step 2}: In the dispersive regime, i.e.,$|g_1/\Delta_1|\ll 1$
and $|g_2/\Delta_2|\ll 1$ (with $\Delta_j=\omega_j-\omega_r$ being
the detuning between the $j$th SCQ and the TLR), and
$\Delta_1=\Delta_2=\Delta$ and $g_1=g_2=g$ (achieved by adjusting
the external bias flux), the efficient Hamiltonian of this system
reads
\begin{eqnarray}
H_{eff}&=&-\frac{g^2}{2\Delta}[4(a^\dagger
a+\frac{1}{2})(\sigma_{z1}+\sigma_{z2})\nonumber\\
&&-(\sigma_{x1}\sigma_{x2}+\sigma_{y1}\sigma_{y2})].
\end{eqnarray}
This Hamiltonian generates a two-qubit gate (with the evolution
duration $t_s=3\pi\Delta/2g^2$)
\begin{eqnarray}
U(t_s)=\exp(-iH_{eff}t_s)=\left(
\begin{array}{cccc}
e^{i\delta} & 0 & 0 & 0 \\
0 & 0 & i & 0\\
0 & i & 0 & 0\\
0 & 0 & 0 & e^{-i\delta}\\
\end{array}\right),\
\end{eqnarray}
where $\delta=6\pi(n+1/2)$ with $n=\langle a^\dagger a\rangle$ being
the mean photon number in the resonator.
If $n$ is known in advance, the global phase factor $\delta$ can be
eliminated by performing single-qubit phase rotations:
$|0\rangle\rightarrow e^{-i\delta/2}|0\rangle$ and
$|1\rangle\rightarrow e^{i\delta/2}|1\rangle$, on each qubit.
Then, the above two-qubit gate reduces to the desirable $i$SWAP gate
\begin{eqnarray}
U_{iSWAP}=\left(
\begin{array}{cccc}
1 & 0 & 0 & 0 \\
0 & 0 & i & 0\\
0 & i & 0 & 0\\
0 & 0 & 0 & 1\\
\end{array}\right).
\end{eqnarray}
Note that no external microwave drive is applied in this step.
With such an {\it i}SWAP gate, the entanglement between two SCQs is
generated, i.e.,
\begin{eqnarray}
|\psi\rangle_{12}\stackrel{U_{{\it
i}SWAP}}{\rightarrow}|\psi'\rangle_{12}=\frac{1}{2}
(|00\rangle-|01\rangle-|10\rangle-|11\rangle)_{12}.
\end{eqnarray}

{\it Step 3}: Apply the unitary operation $R_{y}(3\pi/4)$ to the
first SCQ for generating a desirable Bell state
$|\Phi_-\rangle_{12}$ from the state $|\psi'\rangle_{12}$ prepared
above, i.e.,
\begin{eqnarray}
|\psi'\rangle_{12}\stackrel{R_{y1}(\frac{3\pi}{4})}{\rightarrow}
|\Phi_-\rangle_{12}=\frac{1}{\sqrt{2}}(|00\rangle-|11\rangle)_{12}.
\end{eqnarray}
Here, the unitary operation $R_{y}(3\pi/4)$ is implemented by
$R_{y}(3\pi/4)=R_x(\pi/4)R_z(3\pi/4)R_x(3\pi/4)$. $R_{z}(3\pi/4)$
and $R_{x}(3\pi/4)$ can be realized by applying microwave drives
with the duration $t_3=3\pi
\Delta_a/[2(\Delta_a+g_1^2/\Delta_1)\Delta_a+(2\epsilon
g_1/\Delta_r)^2]$ ($\Delta_a=\omega_1-\omega_d$) and
$t_4=-3\pi\Delta_r/4\epsilon g_1$, respectively.

Similarly, other Bell states in Eq.(1) can be also generated with
the above method.

Customarily, the quantum-state preparation is experimentally
confirmed by quantum-state tomography (see,
e.g.,~\cite{Martinis2009,DiCarlo09,ChowPRA81,FilippPRL102,BianchettiPRA80}),
i.e., reconstructing its density matrix via a series of measurements
on many copies of the prepared state.
However, based on the same logic proposed in Ref.~\cite{wei-prl-ghz}
the Bell state generated above can be simply confirmed by the
following two steps. First, we perform a projective measurement on
the prepared state in the computational basis. The outputs will be
either the logic state $|00\rangle$ or the $|11\rangle$ state with
the same probability (i.e., $P_{00}=P_{11}=1/2$). Note that this is
not the sufficient condition for the confirmation of the Bell state
$|\Phi_-\rangle_{12}$, since a statistical mixture of the logic
states $|00\rangle$ and $|11\rangle$ with the same probability $1/2$
may also result in the same outputs.
Therefore, besides the above projective measurement performed
directly, another projective measurement in the computational basis
is also required after the unitary operation
$R_y(\pi/4)=R_z(\pi/4)R_x(3\pi/4)R_z(3\pi/4)$ on each SCQ. These
operations evolve the prepared Bell state to another Bell state,
i.e.,
\begin{eqnarray}
|\Phi_-\rangle_{12}\stackrel{R_{y1}(\frac{\pi}{4})R_{y2}(\frac{\pi}{4})}{\rightarrow}
|\Psi_+\rangle_{12}=\frac{1}{\sqrt{2}}(|01\rangle+|10\rangle)_{12}.
\end{eqnarray}
This measurement implies that, if the prepared state is the
desirable Bell state $|\Phi_-\rangle_{12}$, then the measured state
should be $|\Psi_+\rangle_{12}$ and thus the output is either
$|01\rangle$ or $|10\rangle$ (with the same probability $1/2$). This
is because the mixture of the states $|00\rangle$ and $|11\rangle$
would not vanish by the quantum interference after the designed
quantum operations.

\section{Joint spectral measurements of two qubits for testing Bell's inequality}

\subsection{Joint spectral measurements of the qubits for confirming the preparation of
Bell states}
Customarily, completely characterizing an unknown quantum state
requires to tomographically reconstruct its density matrix, by a
series of quantum measurements performed on its many copies (see,
e.g.,~\cite{Martinis2009,DiCarlo09,ChowPRA81,FilippPRL102,BianchettiPRA80}).
Here, we will show that the Bell state generated above, e.g.,
$|\Phi_-\rangle_{12}$, can be confirmed by using the joint spectral
measurements of the two qubits in a significantly simpler fashion.
In the dispersive regime of two-qubit circuit QED system, the
transition frequency of each SCQ is far detuned from the coupled
resonator. And in a framework rotating at the drive frequency
$\omega_d$, the efficient Hamiltonian of the whole system is
\begin{eqnarray}
\mathcal{H}&=&(-\Delta_{r}+\Gamma_1\sigma_{z1}+\Gamma_2\sigma_{z2})a^{\dag}a
+\frac{\tilde{\omega}_1}{2}\sigma_{z1}+\frac{\tilde{\omega}_2}{2}\sigma_{z2}\nonumber\\
&&+\epsilon(a^{\dag}+a),
\end{eqnarray}
where $\Gamma_j=g^2_j/\Delta_j$ and
$\tilde{\omega}_j=\omega_j+\Gamma_j$, $j=1,2$. From the first term
in Eq.~(11), it can be directly seen that the resonator is pulled by
different logic states of two qubits. The frequency shifts of the
resonator by $-\Gamma_1-\Gamma_2$, $-\Gamma_1+\Gamma_2$,
$\Gamma_1-\Gamma_2$, and $\Gamma_1+\Gamma_2$ correspond to the logic
states of two qubits $|11\rangle$, $|10\rangle$, $|01\rangle$, and
$|00\rangle$, respectively. Thus, the frequency shifts of the
resonator can be used to mark the logic states of two qubits.

Further, by considering all the statistical quantum correlations
between SCQs and resonator (i.e., beyond the usual
coarse-grained/mean-field approximation), the steady-state
transmission spectra of the driven resonator $S_{ss}=\langle
a^{\dagger}a\rangle_{ss}/2\epsilon$ can be analytically derived
as~\cite{Wei,Huang}
\begin{eqnarray}
S_{ss}=-\frac{2(AC+BD)}{\kappa(A^2+B^2)},
\end{eqnarray}
with
$A=(\Gamma_1^2-\Gamma_2^2)+2(\kappa^2/4-\Delta_r^2)(\Gamma_1^2+\Gamma_2^2)
+(\kappa^2/4-\Delta_r^2)^2-\kappa^2\Delta_r^2$,
$B=-2\kappa\Delta_r(\Gamma_1^2+\Gamma_2^2+\kappa^2/4-\Delta_r^2)$,
$C=\kappa\langle
\sigma_{z1}(0)\sigma_{z2}(0)\rangle_2\Gamma_1\Gamma_2-\kappa\Delta_r(\langle
\sigma_{z1}(0)\rangle_2\Gamma_1+\langle
\sigma_{z2}(0)\rangle_2\Gamma_2)+\kappa/2(3\Delta_r^2-\kappa^2/4-\Gamma_1^2-\Gamma_2^2)$,
and $D=-2\langle
\sigma_{z1}(0)\sigma_{z2}(0)\rangle_2\Delta_r\Gamma_1\Gamma_2-\sum_{j=1}^2\langle
\sigma_{zj}(0)\rangle_j\Gamma_j(\Gamma_j^2-\Gamma_{j'}^2+\kappa^2/4-\Delta_r^2)
+\Delta_r(\Gamma_1^2+\Gamma_2^2+3\kappa^2/4-\Delta_r^2)$, $j\neq
j'=1,2$, $\Gamma_j=g_j^2/\Delta_j$. Here, $\kappa$ denotes photon
leakage rate of the resonator. Experimentally, the qubit decay rate
(e.g., $\gamma=2\pi \times 0.25$MHz~\cite{WallraffPRL05}) is
negligible, since it is significantly smaller than the photon
leakage rate of the resonator (e.g., $\kappa=2\pi \times
1.69$MHz~\cite{BianchettiPRA80}).
It has been numerically proven that, each peak in the above
steady-state spectra marks one of the logic states and the relative
height of such a peak corresponds to the superposed probability in
the detected two-qubit state.

\begin{figure}[htbp]
  \centering
  \includegraphics[width=7cm]{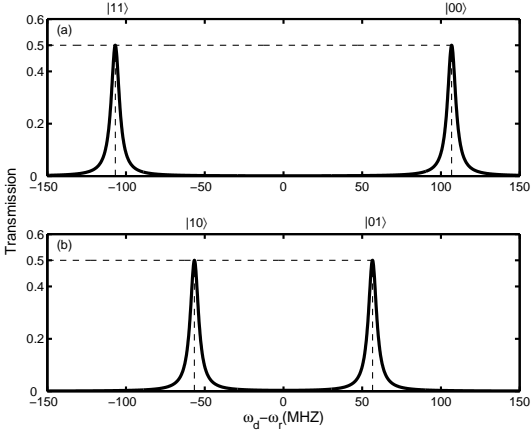}\\
  \caption {Transmission spectra of the driven resonator versus the detuning
  $\Delta_r=\omega_d-\omega_r$ for Bell states $|\Phi_-\rangle_{12}$ (a)
  and $|\Psi_+\rangle_{12}$ (b). The parameters are chosen as $(\Gamma_1, \Gamma_2, \kappa)
  =2\pi \times (13, 4, 1)$MHz~\cite{ChowPRA81}.}\label{Fig3}
\end{figure}

Therefore, if the two SCQs are prepared at the state
$|\Phi_-\rangle_{12}$, then two peaks marking respectively
$|00\rangle$ and $|11\rangle$ with the same height would be detected
in the measured spectra. Furthermore, if the joint spectral
measurements are performed after the evolution (10), then other two
peaks marking respectively $|10\rangle$ and $|01\rangle$ should be
detected in the spectra. This is the direct evidence showing that
the state generated above is just the desirable Bell state, i.e.,
the coherent superposition of the states $|00\rangle$ and
$|11\rangle$, rather than their mixture.
This argument could be verified by the numerical results shown in
Fig.~3. For the typical experimental parameters $(\Gamma_1,
\Gamma_2, \kappa)=2\pi \times (13, 4, 1)$MHz~\cite{ChowPRA81}, the
Fig.~3(a) shows the steady-state spectral distributions (versus the
detuning $\Delta_r=\omega_d-\omega_r$) for the Bell state
$|\Phi_-\rangle_{12}$, and Fig.~3(b) for another Bell state
$|\Psi_+\rangle_{12}$.

\subsection{Testing Bell's inequality via joint spectral measurements}

Being able to measure Bell states constitutes a good entanglement
witness, but we can in fact do better than that. With the generated
Bell state $|\Phi_-\rangle_{12}$, we now show how to test Bell's
inequality with circuit QED via joint spectral measurements of the
distant SCQs.

First, with the single-qubit gates~\cite{Blais07}, a Hadamard-like
operation~\cite{Wei0506} can be constructed as
\begin{eqnarray}
R_j(\theta_j)&=&R_z(\theta_j/2)R_x(\pi/4)R_z(-\theta_j/2)\nonumber\\
&=&\frac{1}{\sqrt{2}}\left(
\begin{array}{cc}
1 & ie^{i\theta_j} \\
ie^{-i\theta_j} & 1 \\
\end{array}\right),\
\end{eqnarray}
which plays an important role for testing Bell's inequality.

Then, the classical variables $\{\theta_1,\theta_2\}$ can be encoded
into the generated Bell state $|\Phi_-\rangle_{12}$ with
Hadamard-like operations $R_j(\theta_j)$. Thus, the state
$|\Phi_-\rangle_{12}$ is changed into
\begin{eqnarray}
|\Phi'_-\rangle_{12}&=&R_1(\theta_1)R_2(\theta_2)|\Phi_-\rangle_{12}\nonumber\\
&=&\frac{1}{2\sqrt{2}}[(1+e^{-i(\theta_1+\theta_2)})|00\rangle
+i(e^{i\theta_2}-e^{-i\theta_1})|01\rangle\nonumber\\
&+&i(e^{i\theta_1}-e^{-i\theta_2})|10\rangle-(1+e^{i(\theta_1+\theta_2)})|11\rangle]_{12}.
\end{eqnarray}

Thirdly, the correlation between two classical variables
$\{\theta_1,\theta_2\}$ is measured by the difference in the
probabilities of two qubits found in the same and different logic
states. This means that the correlation function $E(\theta_1,
\theta_2)$ can be obtained as
\begin{eqnarray}
E(\theta_1, \theta_2)&=&P_{same}(\theta_1,\theta_2)-P_{diff}(\theta_1,\theta_2)\nonumber\\
&=&P_{00}(\theta_1, \theta_2)+P_{11}(\theta_1,
\theta_2)-P_{01}(\theta_1, \theta_2)\nonumber\\
&&-P_{10}(\theta_1,\theta_2),
\end{eqnarray}
where $P_{00}(\theta_1, \theta_2)$\,\,($P_{01}(\theta_1, \theta_2)$,
$P_{10}(\theta_1, \theta_2)$, $P_{11}(\theta_1, \theta_2)$) denotes
the probability of two qubits found in the logic state
$|00\rangle_{12}$\,($|01\rangle_{12}$, $|10\rangle_{12}$,
$|11\rangle_{12}$).

Theoretically, the correlation of two classical variables
$\{\theta_1,\theta_2\}$ can be described by the operator
$T=|11\rangle\langle 11|+|00\rangle\langle 00|-|10\rangle\langle
01|-|01\rangle\langle 01|=\sigma_{z1}\otimes\sigma_{z2}$ and thus
the correlation function is calculated as
\begin{eqnarray}
E(\theta_1, \theta_2)=_{12}\langle
\Phi'_-|T|\Phi'_-\rangle_{12}=\cos(\theta_1+\theta_2).
\end{eqnarray}
Thus, for the set of classical local variables $\{\theta_1,
\theta_2, \theta'_1, \theta'_2\}$, the so-called CHSH
function~\cite{CHSH1969} reads
\begin{eqnarray}
f(|\Phi'_-\rangle_{12})&=&|E(\theta_1, \theta_2)+E(\theta'_1,
\theta_2)+E(\theta_1, \theta'_2)\nonumber\\
&&-E(\theta'_1, \theta'_2)|\nonumber\\
&=&|\cos(\theta_1+\theta_2)+\cos(\theta'_1+\theta_2)\nonumber\\
&&+\cos(\theta_1+\theta'_2)-\cos(\theta'_1+\theta'_2)|.
\end{eqnarray}
Typically, for one set of classical local variables $\{\theta_1,
\theta_2, \theta'_1, \theta'_2\}=\{\pi/4, 3\pi/4, \pi/2, \pi \}$,
the CHSH function~\cite{CHSH1969} is calculated as
\begin{eqnarray}
f(|\Phi'_-\rangle_{12})=\sqrt{2}+1>2.
\end{eqnarray}
Obviously, the CHSH-type Bell's inequality~\cite{CHSH1969} $f\leq2$
is violated.
While, for another set of classical local variables $\{\theta_1,
\theta_2, \theta'_1, \theta'_2\}=\{\pi/4, 0, 7\pi/4, 3\pi/2 \}$, the
CHSH function~\cite{CHSH1969} is
\begin{eqnarray}
f(|\Phi'_-\rangle_{12})=2\sqrt{2}>2.
\end{eqnarray}
Therefore, in this case the CHSH-type Bell's
inequality~\cite{CHSH1969} $f\leq2$ is violated maximally.

The above theoretical predictions can be numerically tested by the
experimental joint spectral measurements. Indeed, the correlation
function in Eq.~(15) can be directly calculated from the measured
transmission spectra of the driven resonator.
\begin{figure}[htbp]
  \centering
  \includegraphics[width=7cm]{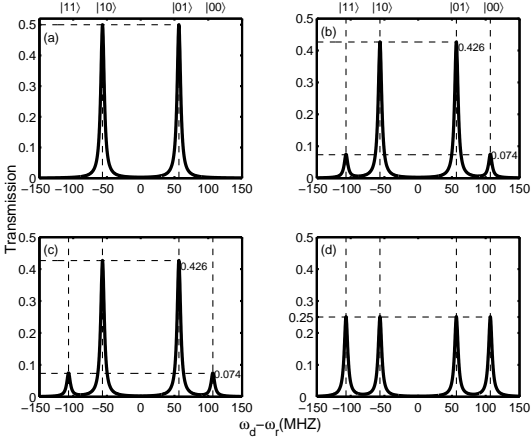}\\
  \caption {Transmission spectra of the driven resonator versus the detuning
  $\Delta_r=\omega_d-\omega_r$ for the state $|\Phi'_-\rangle_{12}$ with
  one set of classical local variables
  $\{\theta_1, \theta_2,\theta'_1, \theta'_2\}=\{\pi/4, 3\pi/4, \pi/2, \pi \}$.
  (a)-(d): the correlation functions can be
  directly calculated, ($E(\theta_1,\theta_2)$,
  $E(\theta'_1,\theta_2)$, $E(\theta_1,\theta'_2)$,
  $E(\theta'_1,\theta'_2))=(-1, -0.704, -0.704, 0)$, according to Eq.~(15).
  The parameters are are the same as that in Fig.~3.}\label{Fig4}
\end{figure}
For example, for the set of classical local variables $\{\theta_1,
\theta_2, \theta'_1, \theta'_2\}=\{\pi/4, 3\pi/4, \pi/2, \pi \}$,
Fig.~4 shows the relevant transmission spectra (corresponding to the
state $|\Phi'_-\rangle_{12}$) versus the detuning of
$\Delta_r=\omega_d-\omega_r$. Here, the parameters are chosen as
$(\Gamma_1, \Gamma_2, \kappa)=2\pi \times (13, 4,
1)$MHz~\cite{ChowPRA81}. With the Fig.~4(a)-(d), the correlation
functions between two classical local variables can be directly
calculated ($E(\theta_1,\theta_2)$, $E(\theta'_1,\theta_2)$,
$E(\theta_1,\theta'_2)$, $E(\theta'_1,\theta'_2))=(-1, -0.704,
-0.704, 0)$. Thus the CHSH function~\cite{CHSH1969} is calculated as
\begin{eqnarray}
f(|\Phi'_-\rangle_{12})=2.408\approx \sqrt{2}+1>2.
\end{eqnarray}
This indicates that the CHSH-type Bell's inequality~\cite{CHSH1969}
$f\leq2$ is violated.
\begin{figure}[htbp]
  \centering
  \includegraphics[width=7cm]{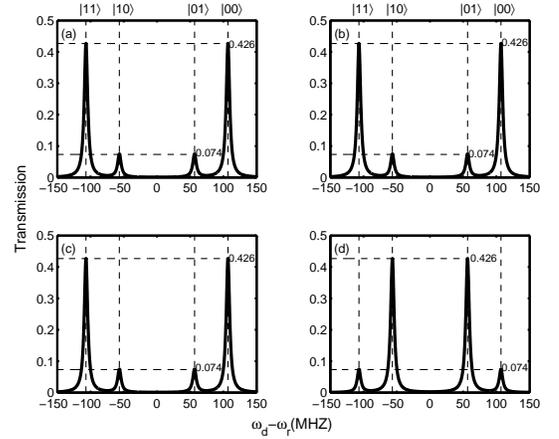}\\
  \caption {Plot of the transmission spectra of the driven resonator versus the detuning
  $\Delta_r=\omega_d-\omega_r$ for the state $|\Phi'_-\rangle_{12}$ with another set of
  classical local variables $\{\theta_1, \theta_2,\theta'_1, \theta'_2\}=\{\pi/4, 0, 7\pi/4, 3\pi/2 \}$.
  (a)-(d): the correlation functions can be directly
  calculated as ($E(\theta_1,\theta_2)$,
  $E(\theta'_1,\theta_2)$, $E(\theta_1,\theta'_2)$,
  $E(\theta'_1,\theta'_2))=(0.704, 0.704, 0.704, -0.704)$.
  The parameters are the same as that in Fig.~3.}\label{Fig5}
\end{figure}
Similarly, for another set of classical local variables $\{\theta_1,
\theta_2, \theta'_1, \theta'_2\}=\{\pi/4, 0, 7\pi/4, 3\pi/2 \}$,
with the same experimental parameters, we plot the transmission
spectra of the driven resonator versus the detuning
$\Delta_r=\omega_d-\omega_r$ for the state $|\Phi'_-\rangle_{12}$ in
Fig.~5. Again, from the Fig.~5(a)-(d) the correlation functions are
directly calculated as ($E(\theta_1,\theta_2)$,
$E(\theta'_1,\theta_2)$, $E(\theta_1,\theta'_2)$,
$E(\theta'_1,\theta'_2))=(0.704, 0.704, 0.704, -0.704)$. As a
consequence, we easily obtain the CHSH function~\cite{CHSH1969}:
\begin{eqnarray}
f(|\Phi'_-\rangle_{12})=2.816\approx 2\sqrt{2}>2.
\end{eqnarray}
Obviously, the CHSH-type Bell's inequality~\cite{CHSH1969} $f\leq2$
is significantly violated.

Finally, we emphasize that compared with the previous schemes for
testing Bell's inequality (see, e.g.,~\cite{Martinis2009,
ChowPRA81}) with superconducting qubits, the present proposal seems
much simpler and easier to be experimentally realized.
We discuss its advantages next.\\

\section{Discussion and Conclusion}

We now briefly address the experimental feasibility of our scheme.
With the typical experimental parameters ($\omega_r, \omega_1,
\omega_2, g_{1(2)}$)=2$\pi \times$(6.442, 4.5, 4.85,
0.133)GHz~\cite{FilippPRL102} and the amplitude of the drive chosen
as $\epsilon$=2$\pi \times$1.2GHz, we can approximately estimate the
time needed in our scheme. The required times for realizing the
unitary operatons $R_x(\pi/4)$, $R_x(3\pi/4)$ (setting
$\omega_d=2\pi\times$4.491GHz to satisfy $\Delta_a+\Gamma_j=0$),
$R_z(3\pi/4)$ (setting $\omega_d=2\pi\times$4GHz), an $i$SWAP gate
(setting $\Delta=2\pi \times1.18$GHz) are $t_{1(2)}$=1.5ns,
$t_4$=4.5ns, $t_3$=1.5ns, and $t_s$=50ns, respectively.
Experimentally, the time interval for performing a joint spectral
measurement introduced above is about 40ns~\cite{BianchettiPRA80}.
Thus, the times for generating a Bell state and confirming its
existence are about $60.5$ns and $93$ns, respectively. Also, the
time interval for testing Bell's inequality can be estimated as
$\sim 160$ns. Finally, the total time in our scheme to perform a
test experiment is about $313.5$ns, which is still shorter than the
qubit's relaxation and dephasing times, e.g., $T_1$=7.3$\mu$s and
$T_2$=500ns~\cite{WallraffPRL05}. It is clear, however, that this is
not enough to close the locality loophole. Therefore, our proposal
should be experimentally realized with the current techniques.

In conclusion, we have proposed a simple and feasible method to test
Bell's inequality with experimentally-demonstrated circuit QED
system by joint spectral measurements. At first, an alternative
method to generate an ideal Bell state is proposed with an $i$SWAP
gate and several single-qubit gates~\cite{Blais07}. Then we present
a simple method to confirm the generation of the Bell state with two
single-qubit unitary operations and two projective measurements on
two qubits. With the experimental joint spectral measurements, the
confirmation can be easily achieved. Further, with two Hadamard-like
operations constructed with single-qubit gates~\cite{Blais07}, two
classical variables $\{\theta_1,\theta_2\}$ can be encoded into the
generated Bell state. For the typical sets of classical local
variables $\{\theta_j,\theta_j', j=1,2\}$, the correlation functions
in Bell's inequality can be directly read out by joint spectral
measurements because each detected peak marks one of the logics
states and its relative height corresponds to the superposed
probability in the detected two-qubit states. In this way, the
Bell's inequality could be efficiently tested. Compared with the
previous schemes for testing Bell's inequality, the advantage of our
proposal is that the desirable coincided measurements on the two
qubits can be directly realized by joint spectral measurements and
the nonlocal correlations between the distant qubits can be directly
read out from the measured transmission spectra of the driven
resonator. As a result, the present proposal seems much simpler and
easier to be experimentally realized by joint spectral measurements.
We believe that our method could be generalized to test Bell-like
inequalities for multi-qubit states in a straightforward way.\\

{\bf Acknowledgements}: This work was supported in part by the
Natural Science Foundation of China under Grant Nos. 10874142 and
90921010, the Fundamental Research Program of China under Grant No.
2010CB923104, and the Fundamental Research Funds for the Central
Universities under Grant Nos. SWJTU09CX078 and 2010XS47. V.V.
acknowledges financial support from the Engineering and Physical
Sciences Research Council in United Kingdom as well as the National
Research Foundation and the Ministry of Education in Singapore.

\end{document}